\documentclass[%
aip,
amsmath,amssymb,
reprint,%
]{revtex4-1}

\usepackage{graphicx}% Include figure files
\usepackage{dcolumn}% Align table columns on decimal point
\usepackage{bm}% bold math
\usepackage[utf8]{inputenc}
\usepackage[T1]{fontenc}
\usepackage{mathptmx}
\usepackage{graphicx}
\usepackage{epstopdf}
\usepackage{amsmath}
\usepackage{diagbox}
\usepackage{color, soul}
\begin{document}
	
	\preprint{AIP/123-QED}
	
	\title[Sample title]{The Similarity of Dynamic behavior between Different Chaotic Systems}
	% Force line breaks with \\
	
\author{Jizhao Liu}
\email{liujizhao@mail.sysu.edu.cn}
\affiliation{School of Data Science and Computer Science, Sun Yat-sen University. No.132, East Outer Ring Road, Guangzhou, Guangdong, P. R. China.\\
}
\author{Xiangzi Zhang}%
\affiliation{School of Information Science and Technology, Jinan University. No.601, West Huangpu Avenue, Guangzhou, Guangdong, P. R. China.\\
}%

\author{Jing Lian}%
\affiliation{School of Electronics and Information Engineering, Lanzhou Jiaotong University\\
	88 West Anning Road, Lanzhou, Gansu 730070, P. R. China\\
}%

\author{Yide Ma}%
\affiliation{School of Information Science and Engineeing, Lanzhou University\\
	Lanzhou, Gansu, China\\
}%

\author{Pengbin Chang}%
\affiliation{School of Information Science and Engineeing, Lanzhou University\\
	Lanzhou, Gansu, China\\
}%

\author{Fangjun Huang}%
\affiliation{School of Data Science and Computer Science, Sun Yat-sen University. No.132, East Outer Ring Road, Guangzhou, Guangdong, P. R. China.\\
}%

\begin{abstract}
	Chaos is associated with stochasticity, complex, irregular motion, etc. It has some peculiar properties such as ergodicity, highly initial value sensitivity, non-periodicity and long-term unpredictability. These pseudo random features lead chaotic systems to enormous applications such as random number generator, image encryption and secure communication. In general, the concept of chaos is never associated with similarity. However, we found the chaotic systems belonging to one chaos family (OCF) have similar dynamic behavior, which is a novel characteristic of chaos.
	\par 
	In this work, three classical chaotic system family are studied, which are Lorenz family, Chua family and hyperbolic sine family. These systems contain different derived chaotic systems (Lorenz system, Chen system and L\"u system), different order chaotic systems (Chua family and hyperbolic sine family), and different kinds of chaotic systems (chaos and hyper-chaos). Their PSPs demonstrate that there exist strong correlation in OCF. Moreover, we found that high order/dimensional chaotic systems will inherit all dynamic behavior of lower ones, and the similarity will decrease as the order/dimensional goes higher, which is analogous to genetic process in biology. All of these features are quantitatively evaluated by PPMCC and SSIM.
\end{abstract}

\maketitle
%\begin{multicols}{2}
\section{Introduction} \label{Sec:Introduction}
Since Lorenz discovered chaos in a simple system of ordinary differential equations in 1959, a new field of science which has grown ever larger with each passing year has been unleashed\cite{sprott2010elegant}. Over decades, chaos has been observed in nature (weather and climate\cite{yuan2018overview}, dynamics of satellites in the solar system\cite{voosen2018meteorite}, time evolution of the magnetic field of celestial bodies\cite{wang2018shadows}, and population growth in ecology\cite{gilpin2017phase}) and laboratory (electrical circuits\cite{qi2017energy}, lasers\cite{zhao2018simultaneous}, chemical reactions\cite{vaidyanathan2017adaptive}, fluid dynamics\cite{zhang2008liquid}, mechanical systems\cite{song2017stabilization}, and magneto-mechanical devices\cite{malaji2018analysis}). Chaotic behavior has also found numerous applications in electrical and engineering\cite{liu2018simplest}, information and communication technologies\cite{wang2018risk}, biology and  medicine\cite{coffey1998self}. By now, many chaotic models have been developed and studied in great detail, but they continue to present surprises and raise questions. The completely features of chaos is not already known.
\par 
There is no universally accepted mathematical definition of chaos. Usually, chaos can often be qualitatively identified with some confidence by observing the strange attractor or chaotic sea in a state space plot or Poincar\'{e} section, or quantitatively identified by Lyapunov exponent\cite{sprott1993strange}. This is mainly due to the features and manifestations of chaos. In detail, the main features of a nonlinear dynamical system, exhibiting deterministic chaos for given values of the parameters, are the following\cite{kocarev2011chaos}:
\par 
\leftmargini=3em
\begin{itemize}
	\item \textbf{Sensitive dependence on initial conditions} - two nearby initial conditions on the attractor or in the chaotic sea is separated by a distance which grows exponentially in time, and leads to long-term unpredictability.
	\item \textbf{Ergodicity} - the trajectory winds around forever, never repeating, and a sufficient long trajectory will almost fill the square for most of the initial conditions. 
	\item \textbf{Strange attractors} - a strange attractor contains an infinite number of points bounded within a definite region of the state space (an attractor with fractal dimension). Moreover, many dynamical systems have multiple coexisting attractors, which can be observed by changing initial conditions\cite{li2018multiple}.
\end{itemize}
\par 
These stochastic-like features lead chaotic systems to enormous cryptographic applications. In general, the concept of chaos is never associated with similarity, and there is no correlation among different chaotic systems (DCS). However, we found that the dynamic behaviors of DCS belonging to one chaos family (OCF) are similar, and thereby suggest there exist correlation among DCS of OCF. To illustrate this phenomenon, we have studied Lorenz family, Chua family, and hyperbolic sine family. First, their phase space plots (PSPs) are numerical calculated by 4th-order Runge-Kutta (RK4) method with fixed step size, then the related PSPs among DCS are selected. At last, Pearson correlation coefficient (PPMCC) and structural similarity (SSIM) are utilized to quantitatively evaluate the similarity of PSPs.
\par
The rest of the paper is organized as follows: In Section 2, the similarity evaluation methods utilized in this paper, i.e., the Pearson correlation coecient (PPMCC) and structural similarity (SSIM) are introduced. In Section 3,  the PSPs of Lorenz family, Chua family, hyperbolic sine family are calculated, and the similarity among these families are quantitatively evaluated. Conclusions and future work are drawn in Section 5. 
\section{Similarity Calculation method}  \label{Sec:Sim method}
\subsection{Pearson correlation coefficient} 
The Pearson correlation coefficient (PPMCC) is one of the most commonly used statistical tools to measure the degree of correlation between two data sets $ X $ and $ Y $\cite{xu2018dependent}:
\par 
\begin{equation} 
PPMCC=\dfrac{\sum_{i=1}^{n}(x_{i}-\bar{x})(y_{i}-\bar{y})}{\sqrt{\sum_{i=1}^{n}(x_{i}-\bar{x})^{2}}\sqrt{\sum_{i=1}^{n}(y_{i}-\bar{y})^{2}}}
\label{Eq:PPMCC}
\end{equation}
Here $ (x_{i}, y_{i}) $ are individual paired samples from the data sets $ X $ and $ Y $, and $ n $ is the total number of pairs; $ \bar{x} $ and $ \bar{y} $ are the mean values of the samples in data sets $ X $ and $ Y $. 
\par 
The degree of PPMCC always lies in the range of $ -1 $ to $ +1 $. PPMCC$ >0 $ implies the two variable have positive correlation, PPMCC $ <0 $ means a negative correlation. When PPMCC is higher than $ 0.5 $ (or lower than -0.5) indicate a strong relationship. PPMCC$ =0 $ implies no relationship.
\subsection{Structural Similarity}
The structural similarity (SSIM) index is used for measuring the similarity between two images\cite{wang2004image}. Basically, SSIM consists of three local comparison functions between two signal $ x $ and $ y $, namely luminance comparison, contrast comparison, and structure comparison, which are computed as follows:
\par 
\begin{equation} 
\begin{aligned} 
I(x,y)&=\frac{2\mu_{x}\mu_{y}+C_{1}}{\mu_{x}^{2}+\mu_{y}^{2}+C_{1}}\\
C(x,y)&=\frac{2\sigma_{x}\sigma_{y}+C_{2}}{\sigma_{x}^{2}+\sigma_{y}^{2}+C_{2}}\\
S(x,y)&=\frac{\sigma_{xy}+C_{3}}{\sigma_{x}\sigma_{y}+C_{3}}
\end{aligned} 
\label{Eq:SSIM}
\end{equation}
\par  
In Eq. (\ref{Eq:SSIM}), $ \mu_{x} $ and $ \mu_{y} $ are the sample means of $ x $ and $ y $, $ \sigma_{x} $ and $ \sigma_{y} $ are the sample standard deviations of $ x $ and $ y $, and $ \sigma_{xy} $ is the sample correlation coefficient between $ x $ and $ y $. The constants $ C_{1} $, $ C_{2} $ and $ C_{3} $ are used to stabilize the algorithm when the denominators approach to zero.
\par 
The general form of SSIM index is given by combining the three comparison functions:
\begin{equation} 
SSIM=l(x,y)^{\alpha} \cdot C(x,y)^{\beta} \cdot S(x,y)^{\gamma}
\label{Eq:SSIM1}
\end{equation} 
where $ \alpha $, $ \beta $ and $ \gamma $ are parameters which define the relative importance of the three components.
\par 
Usually, $ \alpha=\beta=\gamma=1 $. Therefore, SSIM can be rewritten as:
\begin{equation} 
SSIM=\frac{(2\mu_{x}\mu_{y}+C_{1})(2\sigma_{xy}+C_{2})}{(\mu_{x}^{2}+\mu_{y}^{2}+C_{1})(\sigma_{x}^{2}\sigma_{y}^{2}+C_{2})}
\label{Eq:SSIM2}
\end{equation}
\section{The similarity of dynamic behavior among different chaotic systems}
In order to demonstrate the similarity of dynamic behavior among different chaotic systems, Chua family, Chen family and hyperbolic sine are taken as example in this paper. These is because these three chaotic families have been well studied which involve chaos and hyperchaos, different derived and different order/dimensional chaotic systems. 
\par 
First, Chua family is studied, the PSPs indicate the dynamic behavior is similar, and thereby suggest there is strong correlation among this OCF. Then the hyperbolic sine family is studied, the resulting PSPs demonstrate that that high order/dimensional chaotic system inherit all dynamic behavior of lower ones, and the similarity will decrease as the order/dimensional goes higher, which is analogous to genetic process in biology. At last, Lorenz family is studied, the PSPs indicate that this phenomenon is widely exist in different derived chaotic systems (Lorenz system, Chen system and L\"u system), different order chaotic systems, and different kinds of chaotic systems (chaos and hyper-chaos). All of these features have been quantitatively evaluated by PPMCC and SSIM.
\subsection{Chua family case}
The dimensionless state equations of canonical Chua's system are as follows\cite{zhang2014equivalent}:
\begin{equation} \label{Eq:3D Chua}
\begin{aligned} 
\dot{x}&=\alpha (y-x-f(x))\\
\dot{y}&=x-y+z\\
\dot{z}&=\beta z\\
\end{aligned}
\end{equation} 
where 
\begin{equation} \label{Eq:Chua f(x)}
f(x)=m_{1}x+\frac{1}{2}(m_{0}-m_{1})(\left | x+1 \right | - \left | x-1 \right |)
\end{equation}
and $ \alpha =10 $, $ \beta =14.87 $, $ m_{0}=-1.27 $ and $  m_{1}=-0.68 $.
\par 
Forth-order Chua's system is as follows\cite{liu2007attractors}:
\begin{equation} \label{Eq:4D Chua}
\begin{aligned} 
\dot{x}&=\alpha (y-x-f(x))\\
\dot{y}&=x-y+z\\
\dot{z}&=\beta z\\
\dot{u}&=\gamma(z+ru)
\end{aligned}
\end{equation}
Here, $ \alpha =10 $, $ \beta =-14.87 $, $ \gamma=-0.0497 $, $ r=27.3333 $, $ m_{0}=-1.27 $ and $  m_{1}=-0.68 $.
\par 
Fifth-order Chua's system is described as follows\cite{simin2007high}:
\begin{equation} \label{Eq:5D Chua}
\begin{aligned} 
\dot{x}&=a (y-x-f(x))\\
\dot{y}&=b(x-y+z)\\
\dot{z}&=cy\\
\dot{u}&=d(z+r_{1}u)\\
\dot{v}&=e(u-r_{2}v)
\end{aligned}
\end{equation} 
Here, $ a=10 $, $ b=1 $, $ c=-14.87 $, $ d=-0.0497 $, $ e=66.7962 $, $ r_{1}=27.3333 $, $ r2=0.05 $, $ m_{0}=-1.27 $ and $  m_{1}=-0.68 $.
\begin{figure}[!htbp] 
	\centering
	\includegraphics[width=0.9\linewidth]{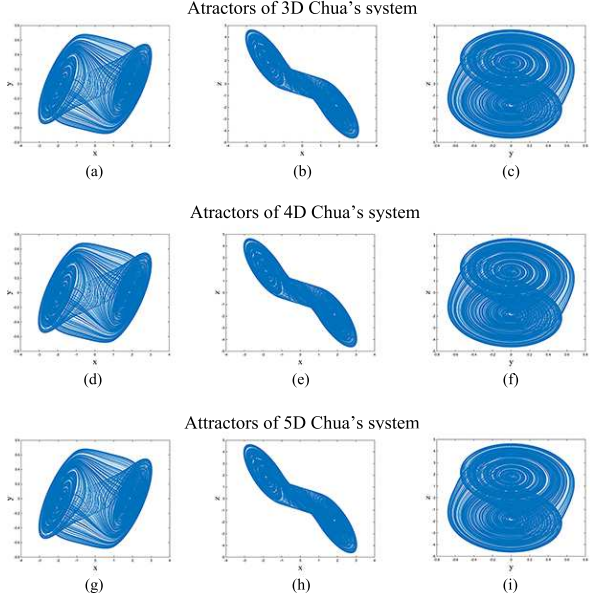}
	\caption{(a) is the phase space plot of $ x-y $ plane of 3D Chua's system, (b) is the phase space plot of $ x-z $ plane of 3D Chua's system, (c) is the phase space plot of $ y-z $ plane of 3D Chua's system, (d) is the phase space plot of $ x-y $ plane of 4D Chua's system, (e) is the phase space plot of $ x-z $ plane of 4D Chua's system, (f) is the phase space plot of $ y-z $ plane of 4D Chua's system, (g) is the phase space plot of $ x-y $ plane of 5D Chua's system, (h) is the phase space plot of $ x-z $ plane of 5D Chua's system, (c) is the phase space plot of $ y-z $ plane of 5D Chua's system}
	\label{fig:nDChua}
\end{figure}
\par
When the step size is set to be $ 0.001 $ and all initial conditions are set to be $ 0.1 $, one can calculate their PSPs as shown Fig. \ref{fig:nDChua}. From Fig. \ref{fig:nDChua}, the corresponding PSPs are almost the same, which indicate the dynamic behavior are very similar to each other.
\par 
\begin{table}[!htb]	 
	\caption{PPMCC score of N-dimensional Chua family}{ 	
		\tiny	
		\newcommand{\tabincell}[2]{\begin{tabular}{@{}#1@{}}#2\end{tabular}}	
		\centering  	
		\begin{tabular}{c|ccccccccc}
			\hline 
			\diagbox & \tabincell{c}{3D system \\ $ x-y $ plane} & \tabincell{c}{3D system \\ $ x-z $ plane} & \tabincell{c}{3D system \\ $ y-z $ plane} & \tabincell{c}{4D system \\ $ x-y $ plane} & \tabincell{c}{4D system \\ $ x-z $ plane} & \tabincell{c}{4D system \\ $ y-z $ plane} & \tabincell{c}{5D system \\ $ x-y $ plane} & \tabincell{c}{5D system \\ $ x-z $ plane} & \tabincell{c}{5D system \\ $ y-z $ plane}\\
			\hline
			\tabincell{c}{3D system \\ $ x-y $ plane}       &  1 & 0.0099  & 0.3588  & \textbf{\underline{0.9531}}  & 0.0099 & 0.3588 & \textbf{\underline{0.9531}} & 0.0099 & 0.3588  \\
			\tabincell{c}{3D system \\ $ x-z $ plane}       &  0.0099 &  1  & 0.0505  & 0.0106  & \textbf{\underline{1}} & 0.0505 & 0.0106 & \textbf{\underline{1}} & 0.0505  \\
			\tabincell{c}{3D system \\ $ y-z $ plane}       &  0.3588 & 0.0505  & 1  & 0.3588  & 0.0505 & \textbf{\underline{1}} & 0.3588 & 0.0505 & \textbf{\underline{1}}  \\
			\tabincell{c}{4D system \\ $ x-y $ plane}       &  \textbf{\underline{0.9531}} & 0.0106  & 0.3588  & 1  & 0.0106 & 0.3588 & \textbf{\underline{1}} & 0.0106 & 0.3588  \\
			\tabincell{c}{4D system \\ $ x-z $ plane}       &  0.0099 & \textbf{\underline{1}} & 0.0505 & 0.0106  & 1 & 0.0505 & 0.0106 & \textbf{\underline{1}} & 0.0505  \\
			\tabincell{c}{4D system \\ $ y-z $ plane}       &  0.3588 & 0.0505  & \textbf{\underline{1}}  & 0.3588  & 0.0505 & 1 & 0.3588 & 0.0505 & \textbf{\underline{1}}  \\
			\tabincell{c}{5D system \\ $ x-y $ plane}       &  \textbf{\underline{0.9531}} & 0.0106  & 0.3588  & \textbf{\underline{1}}  & 0.0106 & 0.3588 & 1 & 0.0106 & 0.3588  \\
			\tabincell{c}{5D system \\ $ x-z $ plane}       &  0.0099 & \textbf{\underline{1}}  & 0.0505  & 0.0106  & \textbf{\underline{1}} & 0.0505 & 0.0106 & 1 & 0.0505  \\
			\tabincell{c}{5D system \\ $ y-z $ plane}       &  0.3588 & 0.0505  & \textbf{\underline{1}}  & 0.3588  & 0.0505 & \textbf{\underline{1}} & 0.3588 & 0.0505 & 1  \\ 
			\hline
	\end{tabular}}
	\label{tab:PPMCC score ND Chua}
\end{table}
\begin{table}[!htb]	 
	\caption{SSIM score of N-dimensional Chua family}{
		\tiny
		\newcommand{\tabincell}[2]{\begin{tabular}{@{}#1@{}}#2\end{tabular}}
		\centering  	
		\begin{tabular}{c|ccccccccc}
			\hline 
			\diagbox & \tabincell{c}{3D system \\ $ x-y $ plane} & \tabincell{c}{3D system \\ $ x-z $ plane} & \tabincell{c}{3D system \\ $ y-z $ plane} & \tabincell{c}{4D system \\ $ x-y $ plane} & \tabincell{c}{4D system \\ $ x-z $ plane} & \tabincell{c}{4D system \\ $ y-z $ plane} & \tabincell{c}{5D system \\ $ x-y $ plane} & \tabincell{c}{5D system \\ $ x-z $ plane} & \tabincell{c}{5D system \\ $ y-z $ plane}\\
			\hline
			\tabincell{c}{3D system \\ $ x-y $ plane}       &  1 & 0.2417  & 0.1949  & \textbf{\underline{0.8776}}  & 0.2417 & 0.1949 & \textbf{\underline{0.8776}} & 0.2417 & 0.1949  \\
			\tabincell{c}{3D system \\ $ x-z $ plane}       &  0.2417 & 1  & 0.1907  & 0.2409  & \textbf{\underline{1}} & 0.1907 & 0.2409 & \textbf{\underline{1}} & 0.1907  \\
			\tabincell{c}{3D system \\ $ y-z $ plane}       &  0.1949 & 0.1907 & 1  & 0.1949  & 0.1907 & \textbf{\underline{1}} & 0.1949 & 0.1907 & \textbf{\underline{1}}   \\
			\tabincell{c}{4D system \\ $ x-y $ plane}       &  \textbf{\underline{0.8776}} & 0.2409  & 0.3588  & 1  & 0.2409 & 0.1949 & \textbf{\underline{1}} & 0.2409 & 0.1949  \\
			\tabincell{c}{4D system \\ $ x-z $ plane}       &  0.2417 & \textbf{\underline{1}} & 0.1907 & 0.0106  & 1 & 0.1907 & 0.2409 & \textbf{\underline{1}} & 0.1907  \\
			\tabincell{c}{4D system \\ $ y-z $ plane}       &  0.1949 & 0.1907  & \textbf{\underline{1}}  & 0.1949  & 0.1907 & 1 & 0.1949 & 0.1907 & \textbf{\underline{1}}  \\
			\tabincell{c}{5D system \\ $ x-y $ plane}       &  \textbf{\underline{0.8776}} & 0.2409  & 0.1949  & \textbf{\underline{1}}  & 0.2409 & 0.1949 & 1 & 0.2409 & 0.1949  \\
			\tabincell{c}{5D system \\ $ x-z $ plane}       &  0.2417 & \textbf{\underline{1}}  & 0.1907  & 0.2409  & \textbf{\underline{1}} & 0.1907 & 0.2409 & 1 & 0.1907  \\
			\tabincell{c}{5D system \\ $ y-z $ plane}       &  0.1949 & 0.1907  & \textbf{\underline{1}}  & 0.1949  & 0.1907 & \textbf{\underline{1}} & 0.1949 & 0.1907 & 1  \\ 
			\hline
	\end{tabular}}
	\label{tab:SSIM score ND Chua}
\end{table}  
\par 
In order to quantitatively describe this features, PPMCC and SSIM are utilized to calculate the similarity of PSPs. The resulting scores are shown in Tab. \ref{tab:PPMCC score ND Chua} and Tab. \ref{tab:SSIM score ND Chua}. From Tab. \ref{tab:PPMCC score ND Chua} and Tab. \ref{tab:SSIM score ND Chua}, some PPMCC and SSIM scores of corresponding PSPs are $ 1 $, which indicate there exist extreme strong correlation among DCS of Chua family.
\subsection{Hyperbolic Sine family case}
Ref\cite{liu2018approach} introduced a a class of simple chaotic systems with hyperbolic sine nonlinearity. With general nth-order ordinary differential equations (ODEs), any desirable order of hyperbolic sine chaotic systems could be constructed. 
\par  
The general form is described by
\begin{equation} \label{Eq:general high-order}
\left\{
\begin{aligned}
&\dot{x_{1}}=x_{2}-x_{1}\\
&\dot{x_{2}}=x_{3}-x_{2}\\
&\cdots\\
&\dot{x_{n-3}}=x_{n-2}-x_{n-3}\\
&\dot{x_{n-2}}=x_{n-1}\\
&\dot{x_{n-1}}=x_{n}\\
&\dot{x_{n}}=-x_{n}-f(x_{n-1})-nx_{n-2}-nx_{n-3}-\cdots-\dfrac{1}{2n}x_{1}
\end{aligned}
\right.
\end{equation} 
where $ n $ is the order of the system. In this system, the nonlinear function is $f(x_{n-1})$, which is defined by $f(x_{n-1})=\rho sinh(\psi\dot{x})$, where $sinh((\psi\dot{x})=(\frac{e^{\psi\dot{x}}-e^{-(\psi\dot{x})}}{2})$, and $\rho=1.2\times 10^{-6}$, $\psi=\frac{1}{0.026}$.
According to this equations, third order hyperbolic sine chaotic system is given by:
\begin{equation} \label{Eq:simplest chaotic system with hyperbolic sine}
\begin{split}
\begin{aligned}
\dot{x_{1}}&=x_{2}\\
\dot{x_{2}}&=x_{3}\\
\dot{x_{3}}&=-x_{3}-\rho sinh(\psi x_{2})-3x_{1}\\
\end{aligned}
\end{split}
\end{equation}
\par 
Fourth order hyperbolic sine chaotic system is given by:
\begin{equation} \label{Eq:fourth order}
\left\{
\begin{aligned}
&\dot{x_{1}}=x_{2}-x_{1}\\
&\dot{x_{2}}=x_{3}\\
&\dot{x_{3}}=x_{4}\\
&\dot{x_{4}}=-x_{4}-\rho sinh(\psi x_{3})-4x_{2}-0.125x_{1}
\end{aligned}
\right.
\end{equation} 
\par 
Fifth order chaotic system is given by:
\begin{equation} \label{Eq:fifth order}
\left\{
\begin{aligned}
&\dot{x_{1}}=x_{2}-x_{1}\\
&\dot{x_{2}}=x_{3}-x_{2}\\
&\dot{x_{3}}=x_{4}\\
&\dot{x_{4}}=x_{5}\\
&\dot{x_{5}}=-x_{5}-\rho sinh(\psi x_{4})-5x_{3}-5x_{2}-0.1x_{1}
\end{aligned}
\right.
\end{equation}
\par  
And tenth order chaotic system is given by:
\begin{equation} \label{Eq:tenth_order}
\left\{
\begin{aligned}
\dot{x_{1}}=&x_{2}-x_{1}\\
\dot{x_{2}}=&x_{3}-x_{2}\\
\dot{x_{3}}=&x_{4}-x_{3}\\
\dot{x_{4}}=&x_{5}-x_{4}\\
\dot{x_{5}}=&x_{6}-x_{5}\\
\dot{x_{6}}=&x_{7}-x_{6}\\
\dot{x_{7}}=&x_{8}-x_{7}\\
\dot{x_{8}}=&x_{9}\\
\dot{x_{9}}=&x_{10}\\
\dot{x_{10}}=&-x_{10}-\rho sinh(\psi x_{9})-10x_{8}-10x_{7}-10x_{6}-10x_{5}-10x_{4}\\
&-10x_{3}-10x_{2}-0.05x_{1}
\end{aligned}
\right.
\end{equation}
\par  
\begin{figure}[!htbp] 
	\centering
	\includegraphics[width=0.9\linewidth]{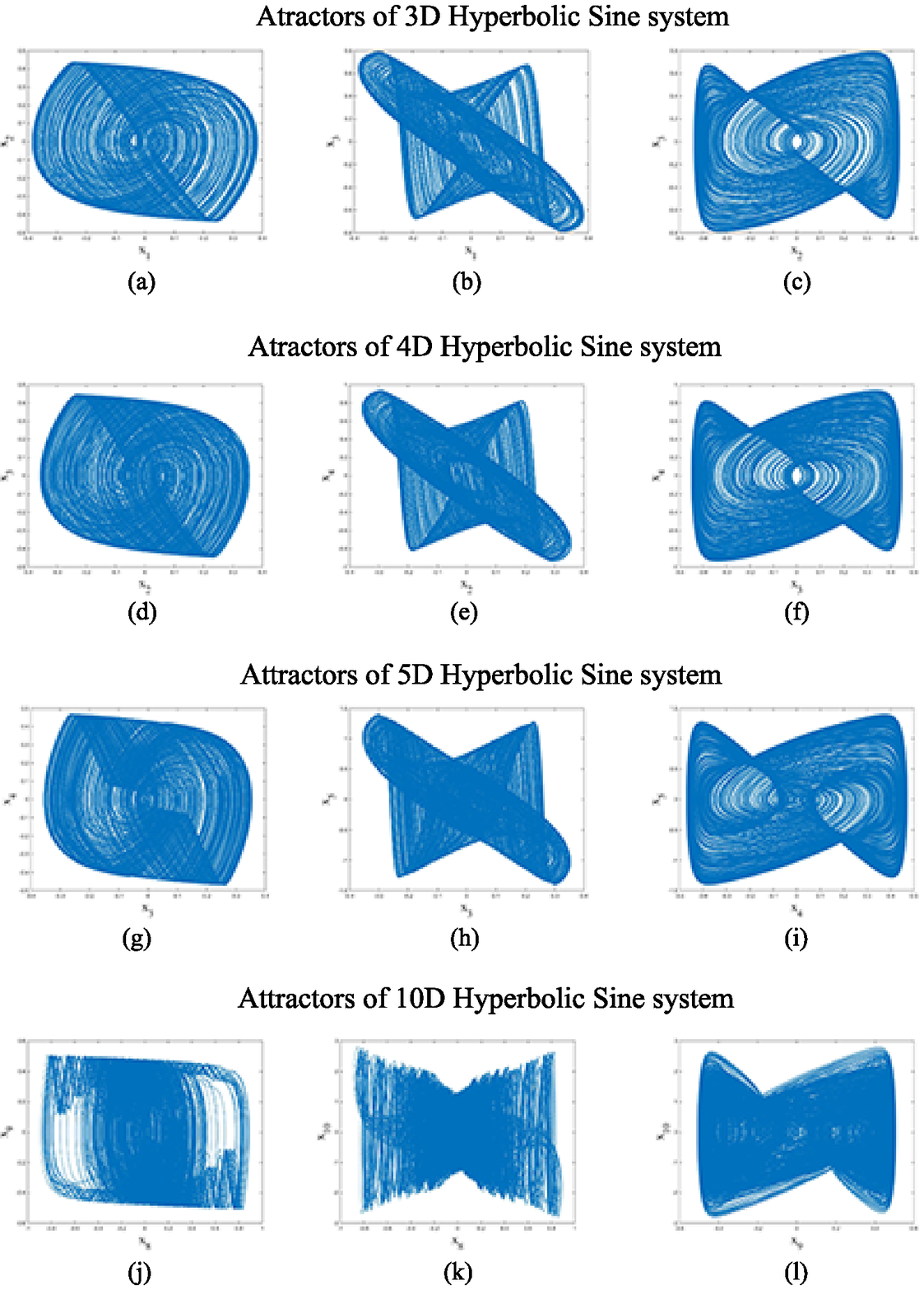}
	\caption{(a) is the phase space plot of $ x_{1}-x_{2} $ plane of 3D hyperbolic sine system, (b) is the phase space plot of $ x_{1}-x_{3} $ plane of 3D hyperbolic sine system, (c) is the phase space plot of $ x_{2}-x_{3} $ plane of 3D hyperbolic sine system, (d) is the phase space plot of $ x_{2}-x_{3} $ plane of 4D hyperbolic sine system, (e) is the phase space plot of $ x_{2}-x_{4} $ plane of 4D hyperbolic sine system, (f) is the phase space plot of $ x_{3}-x_{4} $ plane of 4D hyperbolic sine system, (g) is the phase space plot of $ x_{3}-x_{4} $ plane of 5D hyperbolic sine system, (h) is the phase space plot of $ x_{3}-x_{5} $ plane of 5D hyperbolic sine system, (i) is the phase space plot of $ x_{4}-x_{5} $ plane of 5D hyperbolic sine system, (j) is the phase space plot of $ x_{8}-x_{9} $ plane of 10D hyperbolic sine system, (k) is the phase space plot of $ x_{8}-x_{10} $ plane of 10D hyperbolic sine system, (l) is the phase space plot of $ x_{9}-x_{10} $ plane of 10D hyperbolic sine system}
	\label{fig:ndHS}
\end{figure}
\par
When the step size is set to be $ 0.001 $ and all initial conditions are set to be $ 0.1 $, one can calculate their PSPs as shown Fig. \ref{fig:ndHS}. From the Fig. \ref{fig:ndHS}, the corresponding PSPs are similar, and one can note that the PSPs of lower chaotic system (third order chaotic system) appear in high order chaotic system, which suggest that high order chaotic system inherit all dynamic behavior of lowers ones. However, this similarity is weaken as the orders goes higher.
\par 
\begin{table}[!htb]	 
	\caption{PPMCC score of 3D$\sim$10D hyperbolic sine systems}{ 	
		\tiny	
		\newcommand{\tabincell}[2]{\begin{tabular}{@{}#1@{}}#2\end{tabular}}	
		\centering  	
		\begin{tabular}{c|ccc}
			\hline 
			\diagbox & \tabincell{c}{3D system $ x_{1}-x_{2} $ plane} & \tabincell{c}{3D system $ x_{1}-x_{3} $ plane} & \tabincell{c}{3D system $ x_{2}-x_{3} $ plane}\\
			\hline
			\tabincell{c}{4D system  $ x_{2}-x_{3} $ plane}       &  \textbf{\underline{0.9703}} & 0.3365  & 0.2488 \\
			\tabincell{c}{4D system  $ x_{2}-x_{4} $ plane}       &  0.3635 & \textbf{\underline{0.9458}}  & 0.2208 \\
			\tabincell{c}{4D system  $ x_{3}-x_{4} $ plane}       &  0.2608 & 0.2294  & \textbf{\underline{0.9032}} \\
			\tabincell{c}{5D system  $ x_{3}-x_{4} $ plane}       &  \textbf{\underline{0.8989}} & 0.3452  & 0.2616 \\
			\tabincell{c}{5D system  $ x_{3}-x_{5} $ plane}       &  0.3542 & \textbf{\underline{0.7659}}  & 0.2901 \\
			\tabincell{c}{5D system  $ x_{4}-x_{5} $ plane}       &  0.2488 & 0.2979  & \textbf{\underline{0.7896}} \\
			\tabincell{c}{10D system  $ x_{8}-x_{9} $ plane}       &  \textbf{\underline{0.5789}} & 0.4573  & 0.1673 \\
			\tabincell{c}{10D system  $ x_{8}-x_{10} $ plane}       &  0.4083 & \textbf{\underline{0.5203}}  & 0.4454 \\
			\tabincell{c}{10D system  $ x_{9}-x_{10} $ plane}       &  0.2433 & 0.3635  & \textbf{\underline{0.6765}} \\
			\hline
	\end{tabular}}
	\label{tab:PPMCC score 3D 10D HS}
\end{table}
\begin{table}[!htb]	 
	\caption{SSIM score of 3D$\sim$10D hyperbolic sine systems}{
		\tiny
		\newcommand{\tabincell}[2]{\begin{tabular}{@{}#1@{}}#2\end{tabular}}
		\centering  	
		\begin{tabular}{c|ccc}
			\hline 
			\diagbox & \tabincell{c}{3D system $ x_{1}-x_{2} $ plane} & \tabincell{c}{3D system $ x_{1}-x_{3} $ plane} & \tabincell{c}{3D system $ x_{2}-x_{3} $ plane}\\
			\hline
			\tabincell{c}{4D system  $ x_{2}-x_{3} $ plane}       &  \textbf{\underline{0.7552}} & 0.6211  & 0.5985 \\
			\tabincell{c}{4D system  $ x_{2}-x_{4} $ plane}       &  0.6023 & \textbf{\underline{0.7902}}  & 0.5517 \\
			\tabincell{c}{4D system  $ x_{3}-x_{4} $ plane}       &  0.5583 & 0.5198  & \textbf{\underline{0.6946}} \\
			\tabincell{c}{5D system  $ x_{3}-x_{4} $ plane}       &  \textbf{\underline{0.6987}} & 0.6183  & 0.5993 \\
			\tabincell{c}{5D system  $ x_{3}-x_{5} $ plane}       &  0.6284 & \textbf{\underline{0.7338}}  & 0.5796 \\
			\tabincell{c}{5D system  $ x_{4}-x_{5} $ plane}       &  0.5883 & 0.5805  & \textbf{\underline{0.6813}} \\
			\tabincell{c}{10D system  $ x_{8}-x_{9} $ plane}       &  \textbf{\underline{0.4937}} & 0.4302  & 0.3957 \\
			\tabincell{c}{10D system  $ x_{8}-x_{10} $ plane}       &  0.5062 & \textbf{\underline{0.5416}}  & 0.4508 \\
			\tabincell{c}{10D system  $ x_{9}-x_{10} $ plane}       &  0.5852 & 0.6055  & \textbf{\underline{0.6201}} \\
			\hline
	\end{tabular}}
	\label{tab:SSIM score 3D 10D HS}
\end{table}  
To quantitatively describe this feature, PPMCC and SSIM are utilized to evaluate the similarity scores of PSPs. From Tab. \ref{tab:PPMCC score 3D 10D HS} and \ref{tab:SSIM score 3D 10D HS}, the corresponding PSPs got higher scores than non-corresponding PSPs. However, these scores decrease as order goes higher, which suggested that, high order chaotic system inherit all dynamic behavior of lowers ones, and the correlation is weaken as the orders goes higher. 
\subsection{Lorenz family case}
\subsubsection{Lorenz canonical family}
The equation of canonical Lorenz system is given by\cite{Zhang2011}:
\begin{equation} 
\begin{aligned} 
\dot{x}&=\sigma(y-x)\\
\dot{y}&=-xz+rx-y\\
\dot{z}&=xy-bz
\end{aligned} 
\label{Eq:3D Lorenz}
\end{equation} 
where $ \sigma=10 $, $ r=28 $, and $ b=8/3 $.
\par 
In 1999, Chen found a similar but nonequivalent chaotic attractor, which is now known to be the dual of Lorenz system. The standard form is as follows\cite{chen1999yet}:
\begin{equation} 
\begin{aligned} 
\dot{x}&=a(y-x)\\
\dot{y}&=(c-a)x-xz+cy\\
\dot{z}&=xy-bz
\end{aligned} 
\label{Eq:3D Chen}
\end{equation}
where $ (a,b,c)=(35,3,28)$.
\par  
In 2002, L\"u and Chen reported a new chaotic system which bridging the gap between Lorenz and Chen systems. The equations are given by\cite{lu2002new}:
\begin{equation} 
\begin{aligned} 
\dot{x}&=a(y-x)\\
\dot{y}&=-xz+cy\\
\dot{z}&=xy-bz
\end{aligned}
\label{Eq:3D lv}
\end{equation}
Here, $ a=36 $,  $ b=3 $ and $ c=20 $.
\par  
\begin{figure}[!htbp] 
	\centering
	\includegraphics[width=0.9\linewidth]{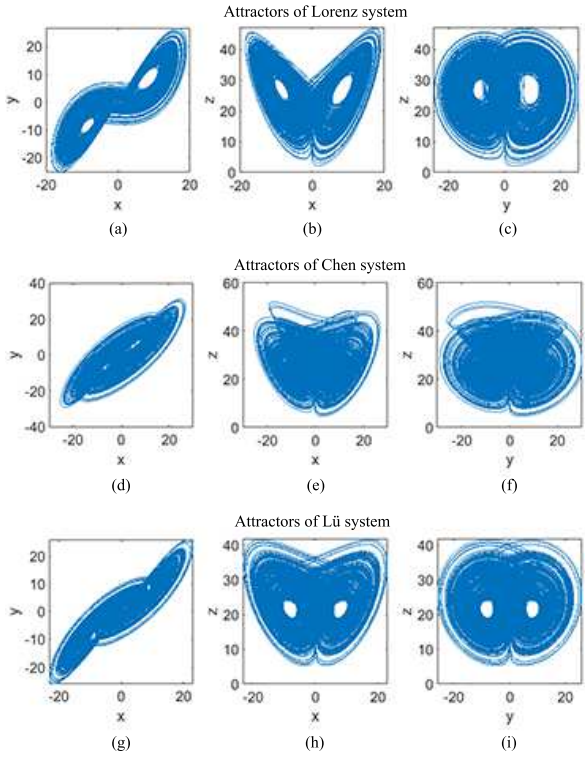}
	\caption{(a)$\scriptsize{\sim}$(c) is phase space plots of Lorenz system, (d)$\scriptsize{\sim}$(f) is phase space plots of Chen system, (g)$\scriptsize{\sim}$(i) is phase space plots of L\"u system.}
	\label{fig:generalLorenz}
\end{figure}
\par 
When the step size is set to be $ 0.001 $ and all initial conditions are set to be $ 0.1 $, one can calculate their PSPs as shown Fig. \ref{fig:generalLorenz}. From Fig. \ref{fig:generalLorenz}, the corresponding PSPs are similar, which indicate there exist correlation among the corresponding dynamic behavior.
\begin{table}[!htb]	
	\caption{PPMCC score of three dimensional Lorenz canonical family}{ 
		\tiny 		
		\newcommand{\tabincell}[2]{\begin{tabular}{@{}#1@{}}#2\end{tabular}}	
		\centering  	
		\begin{tabular}{c|ccccccccc}
			\hline 
			\diagbox & \tabincell{c}{Lorenz system \\ $ x-y $ plane} & \tabincell{c}{Lorenz system \\ $ x-z $ plane} & \tabincell{c}{Lorenz system \\ $ y-z $ plane} & \tabincell{c}{Chen system \\ $ x-y $ plane} & \tabincell{c}{Chen system \\ $ x-z $ plane} & \tabincell{c}{Chen system \\ $ y-z $ plane} & \tabincell{c}{L\"u system \\ $ x-y $ plane} & \tabincell{c}{L\"u system \\ $ x-z $ plane} & \tabincell{c}{L\"u system \\ $ y-z $ plane}\\
			\hline
			\tabincell{c}{Lorenz system \\ $ x-y $ plane}       &  1 & 0.2951  & 0.0521  & \textbf{\underline{0.6851}}  & 0.2002 & 0.1232 & \textbf{\underline{0.6905}} & 0.1593 & 0.0825  \\
			\tabincell{c}{Lorenz system \\ $ x-z $ plane}       &  0.2951 & 1  & 0.3905  & 0.3046  & \textbf{\underline{0.6795}} & 0.3957 & 0.2833 & \textbf{\underline{0.6839}} & 0.4295  \\
			\tabincell{c}{Lorenz system \\ $ y-z $ plane}       &  0.0521 & 0.3905  & 1  & 0.1545  & 0.6204 & \textbf{\underline{0.8260}} & 0.0799 & 0.5877 & \textbf{\underline{0.7532}}  \\
			\tabincell{c}{Chen system \\ $ x-y $ plane}         &  \textbf{\underline{0.6851}} & 0.3046  & 0.1545  & 1  & 0.3098 & 0.2296 & \textbf{\underline{0.7188}} & 0.2872 & 0.2126  \\
			\tabincell{c}{Chen system \\ $ x-z $ plane}         &  0.2002 & \textbf{\underline{0.6795}}  & 0.6204  & 0.3098  & 1 & 0.6555 & 0.2021 & \textbf{\underline{0.8126}} & 0.6644  \\
			\tabincell{c}{Chen system \\ $ y-z $ plane}         &  0.1232 & 0.3957  & \textbf{\underline{0.8260}}  & 0.2296  & 0.6555 & 1 & 0.1402 & 0.5773 & \textbf{\underline{0.7709}}  \\
			\tabincell{c}{L\"u system \\ $ x-y $ plane}         &  \textbf{\underline{0.6905}} & 0.2833  & 0.0799  & \textbf{\underline{0.7188}}  & 0.2021 & 0.1402 & 1 & 0.1669 & 0.0906  \\
			\tabincell{c}{L\"u system \\ $ x-z $ plane}         &  0.1593 & \textbf{\underline{0.6839}}  & 0.5877  & 0.2872  & \textbf{\underline{0.8126}} & 0.5773 & 0.1669 & 1 & 0.7115  \\
			\tabincell{c}{L\"u system \\ $ y-z $ plane}         &  0.0825 & 0.4295  & \textbf{\underline{0.7532}}  & 0.2126  & 0.6644 & \textbf{\underline{0.7709}} & 0.0906 & 0.7115 & 1  \\ 
			\hline
	\end{tabular}} 
	\label{tab:PPMCC score 3D Lorenz}
\end{table}
\begin{table}[!htb]	 
	\caption{SSIM score of three dimensional Lorenz canonical family}{
		\tiny
		\newcommand{\tabincell}[2]{\begin{tabular}{@{}#1@{}}#2\end{tabular}}
		\centering  	
		\begin{tabular}{c|ccccccccc}
			\hline 
			\diagbox & \tabincell{c}{Lorenz system \\ $ x-y $ plane} & \tabincell{c}{Lorenz system \\ $ x-z $ plane} & \tabincell{c}{Lorenz system \\ $ y-z $ plane} & \tabincell{c}{Chen system \\ $ x-y $ plane} & \tabincell{c}{Chen system \\ $ x-z $ plane} & \tabincell{c}{Chen system \\ $ y-z $ plane} & \tabincell{c}{L\"u system \\ $ x-y $ plane} & \tabincell{c}{L\"u system \\ $ x-z $ plane} & \tabincell{c}{L\"u system \\ $ y-z $ plane}\\
			\hline
			\tabincell{c}{Lorenz system \\ $ x-y $ plane}       &   1 & 0.6789  & 0.5602  & \textbf{\underline{0.7694}}  & 0.6074 & 0.5522 & \textbf{\underline{0.7850}} & 0.5507 & 0.5001  \\
			\tabincell{c}{Lorenz system \\ $ x-z $ plane}       & 	0.6789 & 1  & 0.6371  & 0.6583  & \textbf{\underline{0.7064}} & 0.6124 & 0.6515 & \textbf{\underline{0.6730}} & 0.5723  \\
			\tabincell{c}{Lorenz system \\ $ y-z $ plane}       &   0.5602 & 0.6371  & 1  & 0.5659  & 0.6721 & \textbf{\underline{0.8260}} & 0.5343 & 0.6074 & \textbf{\underline{0.6687}}  \\
			\tabincell{c}{Chen system \\ $ x-y $ plane}         &	\textbf{\underline{0.7694}} & 0.3046  & 0.5659  & 1  & 0.6159 & 0.5604 & \textbf{\underline{0.7635}} & 0.5613 & 0.5088  \\
			\tabincell{c}{Chen system \\ $ x-z $ plane}         &	0.6074 & \textbf{\underline{0.7064}}  & 0.6204  & 0.6159  & 1 & 0.6735 & 0.5776 & \textbf{\underline{0.7346}} & 0.6644  \\
			\tabincell{c}{Chen system \\ $ y-z $ plane}         &	0.5522 & 0.6124  & \textbf{\underline{0.7332}}  & 0.5604  & 0.6555 & 1 & 0.1402 & 0.5917 & \textbf{\underline{0.7709}}  \\
			\tabincell{c}{L\"u system \\ $ x-y $ plane}         &	\textbf{\underline{0.7850}} & 0.6515  & 0.5343  & \textbf{\underline{0.7635}}  & 0.5776 & 0.5208 & 1 & 0.5278 & 0.4761  \\
			\tabincell{c}{L\"u system \\ $ x-z $ plane}         &	0.5507 & \textbf{\underline{0.6730}}  & 0.6074  & 0.5613  & \textbf{\underline{0.7346}} & 0.5917 & 0.5278 & 1 & 0.5938  \\
			\tabincell{c}{L\"u system \\ $ y-z $ plane}         &	0.5001 & 0.5723  & \textbf{\underline{0.6687}}  & 0.5088  & 0.6317 & \textbf{\underline{0.6747}} & 0.4761 & 0.5938 & 1  \\ 
			\hline
	\end{tabular}}
	\label{tab:SSIM score 3D Lorenz}
\end{table}
\par 
Tab. \ref{tab:PPMCC score 3D Lorenz} and Tab. \ref{tab:SSIM score 3D Lorenz} are the similarity testing results. One can note that the similarity property exist in different derived chaotic system. Compared to testing results of Chua family and hyperbolic sine family, the scores are lower, which suggest that the dynamic behavior of Chen system and L\"u system are less similar to the dynamic behavior of Lorenz system.
\par 
\subsubsection{High dimensional Lorenz family}
Stenflo derived a fourth order chaotic ordinary differential equations which is known as Lorenz-Stenflo System\cite{zhou1997chaos}. The equations is as follows:
\begin{equation} \label{Eq:4D lorenz}
\begin{aligned} 
\dot{x}&=\sigma(y-x)+sv\\
\dot{y}&=rx-xz-y\\
\dot{z}&=xy-bz\\
\dot{v}&=-x-\sigma v
\end{aligned}
\end{equation} 
Here, $ \sigma=10 $, $ s=28 $, $ r=28 $ and $ b=\frac{8}{3} $.
\par 
The 5D hyperchaootic Lorenz system is given by \cite{yang20135d}:
\begin{equation} \label{Eq:5D lorenz}
\begin{aligned} 
\dot{x}&=\sigma(y-x)+su\\
\dot{y}&=rx-xz-y-v\\
\dot{z}&=xy-\beta z\\
\dot{u}&=k_{1}u-xz\\
\dot{v}&=k_{2}y\\
\end{aligned}
\end{equation} 
where $ \sigma=10 $, $ s=1 $, $ r=28 $, $ \beta=\frac{8}{3} $, $ k_{1}=1.36 $ and $ k_{2}=64 $.
\par
\begin{figure}[!htbp] 
	\centering
	\includegraphics[width=0.9\linewidth]{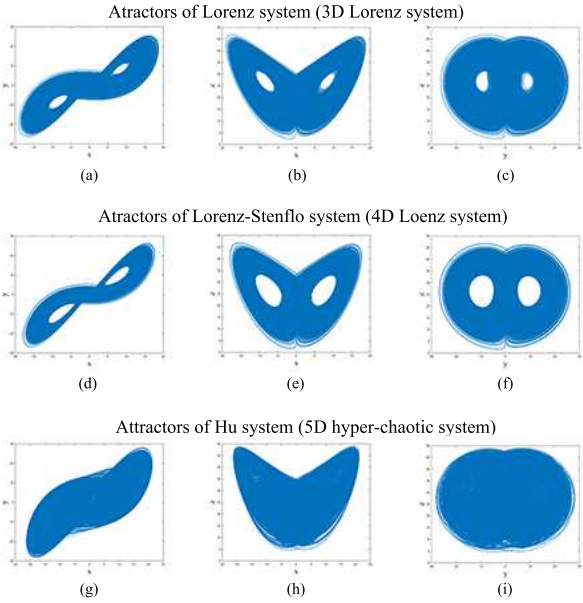}
	\caption{(a) is the phase space plot of $ x-y $ plane of 3D Lorenz system, (b) is the phase space plot of $ x-z $ plane of 3D Lorenz system, (c) is the phase space plot of $ y-z $ plane of 3D Lorenz system, (d) is the phase space plot of $ x-y $ plane of 4D Lorenz system, (e) is the phase space plot of $ x-z $ plane of 4D Lorenz system, (f) is the phase space plot of $ y-z $ plane of 4D Lorenz system, (g) is the phase space plot of $ x-y $ plane of 5D Lorenz system, (h) is the phase space plot of $ x-z $ plane of 5D Lorenz system, (c) is the phase space plot of $ y-z $ plane of 5D Lorenz system,}
	\label{fig:nDLorenz}
\end{figure}

 
When the step size is set to be $ 0.001 $ and all initial conditions are set to be $ 0.1 $, one can calculate their PSPs as shown Fig. \ref{fig:nDLorenz}. From Fig. \ref{fig:nDLorenz}, the corresponding PSPs are similar, which indicate there exist correlation among the corresponding dynamic behavior.
\par
\begin{table}[!htb]	 
	\caption{PPMCC score of N-dimensional Lorenz family}{ 	
		\tiny	
		\newcommand{\tabincell}[2]{\begin{tabular}{@{}#1@{}}#2\end{tabular}}	
		\centering  	
		\begin{tabular}{c|ccccccccc}
			\hline 
			\diagbox & \tabincell{c}{3D system \\ $ x-y $ plane} & \tabincell{c}{3D system \\ $ x-z $ plane} & \tabincell{c}{3D system \\ $ y-z $ plane} & \tabincell{c}{4D system \\ $ x-y $ plane} & \tabincell{c}{4D system \\ $ x-z $ plane} & \tabincell{c}{4D system \\ $ y-z $ plane} & \tabincell{c}{5D system \\ $ x-y $ plane} & \tabincell{c}{5D system \\ $ x-z $ plane} & \tabincell{c}{5D system \\ $ y-z $ plane}\\
			\hline
			\tabincell{c}{3D system \\ $ x-y $ plane}       &  1 & 0.2990  & 0.0709  & \textbf{\underline{0.8181}}  & 0.1722 & 0.0326 & \textbf{\underline{0.7775}} & 0.3444 & 0.1281  \\
			\tabincell{c}{3D system \\ $ x-z $ plane}       &  0.2990 & 1  & 0.4305  & 0.2074  & \textbf{\underline{0.7873}} & 0.3701 & 0.3678 & \textbf{\underline{0.8428}} & 0.4016  \\
			\tabincell{c}{3D system \\ $ y-z $ plane}       &  0.0709 & 0.4305  & 1  & 0.0054  & 0.4856 & \textbf{\underline{0.7968}} & 0.1922 & 0.4120 & \textbf{\underline{0.8072}}  \\
			\tabincell{c}{4D system \\ $ x-y $ plane}       &  \textbf{\underline{0.8181}} & 0.2074  & 0.0054  & 1  & 0.0883 & 0.0731 & \textbf{\underline{0.6308}} & 0.2471 & 0.0557  \\
			\tabincell{c}{4D system \\ $ x-z $ plane}       &  0.1722 & \textbf{\underline{0.7873}}  & 0.4856  & 0.0883  & 1 & 0.5132 & 0.2935 & \textbf{\underline{0.7458}} & 0.3927  \\
			\tabincell{c}{4D system \\ $ y-z $ plane}       &  0.0326 & 0.3701  & \textbf{\underline{0.7968}}  & 0.0731  & 0.5132 & 1 & 0.0753 & 0.3211 & \textbf{\underline{0.6965}}  \\
			\tabincell{c}{5D system \\ $ x-y $ plane}       &  \textbf{\underline{0.7775}} & 0.3678  & 0.1922  & \textbf{\underline{0.6308}}  & 0.2935 & 0.0753 & 1 & 0.4372 & 0.2277  \\
			\tabincell{c}{5D system \\ $ x-z $ plane}       &  0.3444 & \textbf{\underline{0.8428}}  & 0.4120  & 0.2471  & \textbf{\underline{0.7458}} & 0.3211 & 0.4372 & 1 & 0.4308  \\
			\tabincell{c}{5D system \\ $ y-z $ plane}       &  0.1281 & 0.4016  & \textbf{\underline{0.8072}}  & 0.0557  & 0.3927 & \textbf{\underline{0.6965}} & 0.2277 & 0.4308 & 1  \\ 
			\hline
	\end{tabular}}
	\label{tab:PPMCC score ND Lorenz}
\end{table}
\begin{table}[!htb]	 
	\caption{SSIM score of N-dimensional Lorenz family}{
		\tiny
		\newcommand{\tabincell}[2]{\begin{tabular}{@{}#1@{}}#2\end{tabular}}
		\centering  	
		\begin{tabular}{c|ccccccccc}
			\hline 
			\diagbox & \tabincell{c}{3D system \\ $ x-y $ plane} & \tabincell{c}{3D system \\ $ x-z $ plane} & \tabincell{c}{3D system \\ $ y-z $ plane} & \tabincell{c}{4D system \\ $ x-y $ plane} & \tabincell{c}{4D system \\ $ x-z $ plane} & \tabincell{c}{4D system \\ $ y-z $ plane} & \tabincell{c}{5D system \\ $ x-y $ plane} & \tabincell{c}{5D system \\ $ x-z $ plane} & \tabincell{c}{5D system \\ $ y-z $ plane}\\
			\hline
			\tabincell{c}{3D system \\ $ x-y $ plane}       &  1 & 0.6268  & 0.4856  & \textbf{\underline{0.8027}}  & 0.5847 & 0.4665 & \textbf{\underline{0.7585}} & 0.6084 & 0.4848  \\
			\tabincell{c}{3D system \\ $ x-z $ plane}       &  0.6268 & 1  & 0.5494  & 0.6064  & \textbf{\underline{0.7095}} & 0.3701 & 0.6291 & \textbf{\underline{0.7125}} & 0.5287  \\
			\tabincell{c}{3D system \\ $ y-z $ plane}       &  0.4856 & 0.5494  & 1  & 0.4696  & 0.5388 & \textbf{\underline{0.6244}} & 0.5047 & 0.5281 & \textbf{\underline{0.6120}}  \\
			\tabincell{c}{4D system \\ $ x-y $ plane}       &  \textbf{\underline{0.8027}} & 0.2074  & 0.4696  & 1  & 0.5690 & 0.4562 & \textbf{\underline{0.7263}} & 0.5841 & 0.4651  \\
			\tabincell{c}{4D system \\ $ x-z $ plane}       &  0.5847 & \textbf{\underline{0.7095}}  & 0.5388  & 0.5690  & 1 & 0.5406 & 0.6011 & \textbf{\underline{0.6788}} & 0.5060  \\
			\tabincell{c}{4D system \\ $ y-z $ plane}       &  0.4665 & 0.5238  & \textbf{\underline{0.6244}}  & 0.4562  & 0.5406 & 1 & 0.4780 & 0.5046 & \textbf{\underline{0.5826}}  \\
			\tabincell{c}{5D system \\ $ x-y $ plane}       &  \textbf{\underline{0.7585}} & 0.6291  & 0.5047  & \textbf{\underline{0.7263}}  & 0.6011 & 0.4780 & 1 & 0.6351 & 0.5080  \\
			\tabincell{c}{5D system \\ $ x-z $ plane}       &  0.6084 & \textbf{\underline{0.7125}}  & 0.5281  & 0.5841  & \textbf{\underline{0.6788}} & 0.5046 & 0.6351 & 1 & 0.5365  \\
			\tabincell{c}{5D system \\ $ y-z $ plane}       &  0.4848 & 0.5287  & \textbf{\underline{0.6120}}  & 0.4651  & 0.5060 & \textbf{\underline{0.5826}} & 0.5080 & 0.5365 & 1  \\ 
			\hline
	\end{tabular}}
	\label{tab:SSIM score ND Lorenz}
\end{table}
Tab. \ref{tab:PPMCC score ND Lorenz} and Tab. \ref{tab:SSIM score ND Lorenz} are the testing results of PPMCC and SSIM. One can note that, the similarity is also exits in different kinds of chaotic systems (chaos and hyper-chaos). Compared to testing results of Chua family and hyperbolic sine family, the scores are lower, which suggest that the dynamic behavior among hyper-chaotic systems are less similar to the dynamic behavior among high-order chaotic systems. 
\section{Conclusion and future work}
Chaos is famous for its stochastic-like properties such as ergodicity, highly initial value sensitivity, non-periodicity and long-term unpredictability. These pseudo random features lead chaotic systems to enormous applications such as random number generator, image encryption and secure communication. 
\par 
In this paper, we found the dynamic behavior among DCS belonging to OCF is analogous to genetic process in biology, that is: 1. The dynamic behavior is similar in OCF. 2. High order/dimensional chaotic systems will inherit all dynamic behavior of lowers ones, and the similarity will decrease as the orders/dimensional goes higher. 3. The dynamic behavior of different derived chaotic systems and different types of chaotic system are less similar than the dynamic behavior among different orders of chaotic system. This phenomenon is widely exist in derived chaotic systems (Chen system, L\"u system and Lorenz system), different order of chaotic systems (Chua system and hyperbolic sine chaotic system), and different dimensional chaotic systems/hyper-chaotic systems (N-dimensional Lorenz family). These novel features has been quantitatively evaluated by PPMCC and SSIM.
\par 
There is some work which is worth to study in future.
\par 
1. The results need to verify by rigorous mathematical proof. 
\par 
2. The method to evaluate the similarity among different chaotic systems should be improved.
\section*{Acknowledgment}
Conflict of Interest: The authors declare that they have no conflict of interest.
	
	\nocite{*}
	\bibliography{ref}% Produces the bibliography via BibTeX.
	
\end{document}